# Higher Order Antibunching and Subpoissonian Photon Statistics in Five Wave Mixing Process

AMIT VERMA AND ANIRBAN PATHAK
amit.verma@jiit.ac.in, anirbanpathak@yahoo.co.in
Department of Physics and Material Science, JIIT University, A-10, Sector-62, Noida, UP-201307, India.

**Introduction:**

The idea of HOA was introduced by Lee in a pioneering paper[1] in 1990, since then it is predicted only in two photon coherent state[1], trio coherent state[2], the interaction of intense laser beam with an inversion symmetric third order nonlinear medium[3], some simple optical processes[4] and intermediate states[5]. Several criteria of HOA have also been proposed[1-3] which can be shown as equivalent[3]. But still HOA appears to be a very rare phenomenon. Recently, Prakash and Mishra[6] have established a generalized criterion for HOSPS but till now neither the relation of HOSPS with HOA nor the possibilities of existence of HOSPS in a real physical system have been studied. Keeping these facts in mind, the present study shows that HOSPS and HOA can be seen in five wave mixing and third harmonic generation processes and at least, in these two processes they appear simultaneously. The condition for $l^{th}$ order antibunching[3] is

$$d(l) = \langle N_x^{(l+1)} \rangle - \langle N_x \rangle^{l+1} < 0 \qquad \ldots\ldots\ldots\ (1)$$

where $N$ is the usual number operator, $\langle N_x^{(i)} \rangle = \langle N(N-1)\ldots(N-i-1) \rangle$ is the $i^{th}$ factorial moment of number operator, $\langle \ \rangle$ denotes the quantum average. Again, Mishra-Prakash[4] condition for $(l-1)^{th}$ order HOSPS is

$$D(l-1) = \sum_{k=0}^{l}\sum_{i=0}^{l-k} {}^lC_k (-1)^k S_2(l-k,i) \langle N^{(i)} \rangle \langle N \rangle^k - \sum_{k=0}^{l}\sum_{i=0}^{l-k} {}^lC_k (-1)^k S_2(l-k,i) \langle N \rangle^{k+i} < 0 \ \ldots \ (2)$$

where $S_2(l,k)$ denotes the Stirling number of second kind. As the HOSPS can be seen from $l=3$, the simplest condition to check the existence of HOSPS is that of second order subpoissonian photon statistics and is given as

$$D(2) = \langle N^{(3)} \rangle + 2\langle N \rangle^3 - 3\langle N^{(2)} \rangle \langle N \rangle + 3\langle N^{(2)} \rangle - 3\langle N \rangle^2 < 0 \qquad \ldots\ldots\ldots\ (3)$$

In section 2, we present a second order operator solution of the equation of motion of five wave mixing process and use that and equations (1) and (3) to show the existence of HOA and HOSPS in five wave mixing process. In section 3 we have studied the possibilities of observing HOA and HOSPS in third harmonic generation process. Finally section 4 is dedicated to conclusions.

**Five wave mixing process:**

Five wave mixing may happen in different ways. One way is that three photon of frequency $\omega_1$ are absorbed (as pump photon) and two photon of frequency $\omega_2$ are emitted. The Hamiltonian representing this particular five wave mixing process is

$$H = A^\dagger A \omega_1 + B^\dagger B \omega_2 + g(A^{\dagger 3}B^2 + A^3 B^{\dagger 2}) \qquad \ldots\ldots\ldots\ (4)$$

where $\hbar = 1$, $A = ae^{i\omega_1 t}$, $B = be^{i\omega_2 t}$ and $a, a^\dagger$ are annihilation, creation operators in pump mode which satisfy $[a,a^\dagger]=1$, similarly $b$ and $b^\dagger$ are annihilation and creation operators in signal mode and $g$ is the coupling constant. Using short time approximation method [for details of the method see[4]], we can solve the equation of motion corresponding to the pump mode of Hamiltonian (4) as

$$A(t) = A - 3igtA^{\dagger 2}B^2 + \frac{3}{2}g^2t^2[6A^\dagger A^2 B^{\dagger 2}B^2 + 6AB^{\dagger 2}B^2 - 4A^{\dagger 2}A^3 B^\dagger B - 2A^{\dagger 2}A^3]. \ \ldots\ldots\ (5)$$

We can now use (5) to check whether it satisfies (1) and (3) or not. To do so we have first calculated the second order analytic form of time evolution of the operators, $N_A(t)=A^\dagger(t)A(t)$, $N_A^{(2)}(t), N_A^{(3)}(t)$ etc. and then we have taken expectations with respect to the state $|\alpha,0\rangle$. This process yields,

$$\langle N_A(t)\rangle = |\alpha|^2 - 6g^2t^2|\alpha|^6 \quad\quad\quad (6)$$

$$\langle N_A^{(2)}(t)\rangle = |\alpha|^4 - 12g^2t^2(|\alpha|^8 + |\alpha|^6) \quad\quad\quad (7)$$

$$\langle N_A^{(3)}(t)\rangle = |\alpha|^6 - 6g^2t^2(3|\alpha|^{10} + 6|\alpha|^8 + 2|\alpha|^6) \quad\quad\quad (8)$$

Where $A|\alpha\rangle = \alpha|\alpha\rangle$. Now by using (6)-(8) we can obtain

$$d(1) = \langle N_A^{(2)}(t)\rangle - \langle N_A(t)\rangle^2 = -12g^2t^2|\alpha|^6$$

$$d(2) = \langle N_A^{(3)}(t)\rangle - \langle N_A(t)\rangle^3 = -12g^2t^2(3|\alpha|^8 + |\alpha|^6) \quad\quad\quad (9)$$

and $\quad\quad D(2) = -48\, g^2 t^2 |\alpha|^6 \quad\quad\quad (10)$

From last two equations it is clear that $d(1)$, $d(2)$ and $D(2)$ are always negative which shows normal antibunching, HOA and HOSPS respectively in five wave mixing process.

**Third harmonic generation:**

The similar procedure can be repeated for third harmonic generation process whose Hamiltonian can be written as

$$H = A^\dagger A \omega_1 + B^\dagger B \omega_2 + g(A^{\dagger 3}B + A^3 B^\dagger) \quad\quad\quad (11)$$

By applying same method as discussed in section 2, we obtain

$$d(1) = -6g^2t^2|\alpha|^6,$$

$$d(2) = -6g^2t^2(3|\alpha|^8 + |\alpha|^6) \quad\quad\quad (12)$$

and $\quad\quad D(2) = -24\, g^2 t^2 |\alpha|^6 \quad\quad\quad (13)$

Equations (12), (13) satisfies both the criteria (1) and (2) for HOA and HOSPS respectively.

**Conclusions:**

Criterion of HOA and HOSPS are satisfied for both the physical systems. The present work along with the observations[4] strongly establishes the fact that HOA can be observed for different modes of multiwave mixing processes. Further we have observed that (the detail of the calculation is not shown here) HOA is not observed in signal mode of the above cases. In general, HOA is not observed for the mode, whose power in the interaction term is lesser. It is also observed that if we assume that the anharmonic constant and number of photon initially present in the pump mode are same for both the cases then the depth of nonclassicality is more in five wave mixing process than that in the third harmonic generation process.

**Acknowledgement:** AP thanks DST for partial financial support through the project no. SR\FTP\PS-13\2004.

**References:**

1. Lee C T, Phys. Rev. A **41** 1721 (1990).
2. An N B, J. Opt. B: Quantum Semiclass. Opt. **4** (2002) 222-227.
3. Pathak A and Garcia M, Applied Physics B **84** (2006) 484
4. Gupta P, Pandey P and Pathak A, J. Phys. B **39** (2006) 1137.
5. Verma A, Sharma N K and Pathak A, quant-ph\0706.0697.
6. H Prakash and D K Mishra, J. Phys. B, **39** (2006) 2291.